\documentclass[RNAAS]{aastex631}

\newcommand{\jzeronine}{4FGL J0935.3+0901}
\newcommand{\jsixteen}{4FGL J1627.7+3219}
\newcommand{\jtwentytwo}{4FGL J2212.4+0708}
\newcommand{\presto}{\texttt{PRESTO}}

\usepackage{booktabs}
\usepackage[caption=false]{subfig}

\begin{document}

\title{No MSP Counterparts Detected in GBT Searches of Spider Candidates\\4FGL J0935.3+0901, 4FGL J1627.7+3219, and 4FGL J2212.4+0708}

\correspondingauthor{Kyle A. Corcoran}
\email{kac8aj@virginia.edu}

\author[0000-0002-2764-7248]{Kyle A. Corcoran}
\affiliation{Department of Astronomy, University of Virginia}

\author[0000-0001-5799-9714]{Scott M. Ransom}
\affiliation{National Radio Astronomy Observatory}

\author[0000-0001-5229-7430]{Ryan S. Lynch}
\affiliation{Green Bank Observatory}



\begin{abstract}

We performed radio searches for the ``spider'' millisecond pulsar (MSP) candidates \jzeronine, \jsixteen, and \jtwentytwo\ using the Green Bank Telescope in an attempt to detect the proposed radio counterpart of the multi-wavelength variability seen in each system.  We observed using the VEGAS spectrometer, centered predominantly at 2165 MHz; however, we were also granted observations at 820\,MHz for \jsixteen.  We performed acceleration searches on each dataset using \presto\ as well as additional jerk searches of select observations.  We see no evidence of a radio counterpart in any of the observations for each of the three systems at this time.  Additional observations, perhaps at different orbital phases (e.g., inferior conjunction), may yield detections of an MSP in the future.  Therefore, we urge continued monitoring of these systems to fully characterize the radio nature, however faint or variable, of each system. 

\end{abstract}



\section{Introduction} \label{sec:intro}

Redback (RB) and black widow (BW) pulsars -- deemed ``spiders'' -- are compact binary systems containing a millisecond pulsar (MSP) and a low-mass companion \citep[$M\gtrsim 0.1 M_{\odot}$ for RBs and $M<0.1M_{\odot}$ for BWs;][]{Roberts2013}.  
These systems are unique among MSPs in that they can be identified through multi-wavelength observations spanning much of the electromagnetic spectrum.  Significant $\gamma$-rays, X-ray photon detection or even light curve modulation, and optical variability have all been used to study the nature of several spider systems as well as to identify candidate systems.

One such candidate system was identified by \citet{Wang2020}.  The candidate RB (and transitional MSP) \jzeronine\ was identified as an unassociated source by \textit{Fermi} LAT (Large Area Telescope), and the system was found to exhibit $\gamma$-ray modulation.  A possible X-ray source was also found in archival data from \textit{Swift} within the \textit{Fermi} uncertainty region for the object.  Archival and follow-up optical observations suggest that the system potentially has an orbital period of $\sim2.5$\,hr as well as spectral features consistent with an accretion disk.  \citet{Wang2020} suggested that follow-up observations be conducted to determine if the binary is in fact a transitional MSP in an accretion-driven state or, potentially, the rotationally-powered MSP state that would follow.

\citet{Braglia2020} also reported two new spider candidates in a similar fashion.  By searching for optical variations in archival survey data for systems that had possible associated X-ray counterparts within the \textit{Fermi} uncertainty region, they identified \jsixteen\ and \jtwentytwo\ as candidates.  

We performed radio searches for each of these candidates with the Green Bank Telescope (GBT) using the VEGAS spectrometer.  In the section that follows, we outline the procedures we followed in analyzing each GBT observation. After, we discuss the implications of our null results and comment on the need for additional observations in the future.

\newpage
\begin{table}
    \centering
    \caption{Summary of our observations on the three spider candidates.}
    \scriptsize
    \begin{tabular}{@{} cccccccccc @{}}
    \toprule
    
    \multicolumn{1}{c}{\scriptsize{4FGL}} &
    \multicolumn{1}{c}{\scriptsize{$P_{\rm orb}$}} &
    \multicolumn{1}{c}{\scriptsize{MJD}} &
    \multicolumn{1}{c}{\scriptsize{Obs. Length}} &
    \multicolumn{1}{c}{\scriptsize{Sample Time}} &
    % \multicolumn{1}{c}{\scriptsize{Coherent Dedisp. DM}} &
    \multicolumn{1}{c}{\scriptsize{DM$^{\dagger}$}} &
    \multicolumn{1}{c}{\scriptsize{$N_{\rm channels}$}} &
    \multicolumn{1}{c}{\scriptsize{Central Freq.}} &
    \multicolumn{1}{c}{\scriptsize{Channel Width}} &
    \multicolumn{1}{c}{\scriptsize{Effective BW}}
    \\
    
    \multicolumn{1}{c}{\scriptsize{}} &
    \multicolumn{1}{c}{\scriptsize{[d]}} &
    \multicolumn{1}{c}{\scriptsize{}} &
    \multicolumn{1}{c}{\scriptsize{[s]}} &
    \multicolumn{1}{c}{\scriptsize{[$\mu$s]}} &
    \multicolumn{1}{c}{\scriptsize{[pc\,cm$^{-3}$]}} &
    \multicolumn{1}{c}{\scriptsize{}} &
    \multicolumn{1}{c}{\scriptsize{[MHz]}} &
    \multicolumn{1}{c}{\scriptsize{[MHz]}} &
    \multicolumn{1}{c}{\scriptsize{[MHz]}}
    
    \\
    
    \midrule
    
    J0935.3+0901 & 0.10292$^{a}$ & 59613 & 2687 & 43.69 & 31.2 & 768 & 2165 & 1.4648 & 900 \\
    &  & 59614 & 907 & & & & & & 750  \\
    \midrule
    %%%%%%%%%%%%%%%%%%%%%%%%%%%%%%%%%%%%%%%%%%%%%%%%%%%%%%%%%%%
    J1627.7+3219 & 0.49927$^{b}$ & 59619 & 2100 & 43.69 & 0.715 & 768 & 2165 & 1.4648 & 850 \\
     & & 59620  & 2452 & 40.96 &  & 2048 & 820 & 0.09765 & 135 \\
    \midrule
    %%%%%%%%%%%%%%%%%%%%%%%%%%%%%%%%%%%%%%%%%%%%%%%%%%%%%%%%%%%
    J2212.4+0708 & 0.31884$^{b}$ & 59613 & 1380 & 43.69 & 41.1 & 768 & 2165 & 1.4648 & 625 \\

    \bottomrule
    
    % \multicolumn{3}{@{}l}{\textbf{\small{Notes}}}\\
    \multicolumn{10}{@{}l}{\scriptsize{$^{a}$ \citep{Wang2020}}} \\
    \multicolumn{10}{@{}l}{\scriptsize{$^{b}$ \citep{Braglia2020}}} \\
    \multicolumn{10}{@{}l}{\scriptsize{$^{\dagger}$ Predicted DM \citep[NE2001;][]{Cordes2002}, based on distance via \citet{Gaia2018} parallax, used for coherent dedispersion}}\\
    
    \end{tabular}
    
    \label{tab:obs}
\end{table}
\section{Observations and Search Procedures} \label{sec:obs&analysis}
We used a 1500\,MHz-bandwidth mode on VEGAS with coherent dedispersion, centered at 2165\,MHz to observe all systems.
%the new 1500\,MHz (S-band) mode on the VEGAS spectrometer centered at 2165\,MHz with coherent dedispersion to observe all systems
This mode allows us to cover roughly 1700--2700\,MHz with an effective bandwidth of almost 1000\,MHz when significant RFI is not present.  We were also granted observations at 820 MHz for \jsixteen, which were taken in the incoherent ``search'' mode.  The S-band data for each observation were combined together (i.e., had the frequency resolution reduced and were converted to total intensity) using the \texttt{psrfits\_subband} routine, and channels containing prominent RFI -- as found via \presto's \texttt{rfifind} routine -- were removed from all observations.  A summary of all observations can be found in Table \ref{tab:obs}.

We performed acceleration searches on each dataset using standard routines in \presto.  
%In the case of \jzeronine, we performed an independent search on each of the two S-band pointings.  Similarly, the S-band and 820\,MHz data for \jsixteen\ were searched independently of each other.  
As not all of the few thousand candidates generated for each observation are necessarily sensible, we took the top 100 candidate parameter sets and folded the time-series data on each candidate.  These preliminary folds allow us to visually inspect for additional RFI as well as candidates with no potential signals in the time-series.  We then folded the remaining candidates on the full observational dataset, which allows us to see if a candidate peaks in significance at a non-zero DM, and we once again visually inspected these plots to further vet each candidate.  Legitimate detections -- or even candidate detections that would need confirmation from an additional observation -- would likely show statistically significant ($\gtrsim5\sigma$) signal throughout the whole observation or some small fraction thereof at the least. We did not find any candidates that we believe are pulsar signals. 
%We did not find any parameter sets that we claim as a detection or candidate detection.

For binaries with short periods, and especially with shorter duration scans, jerk searches can be performed to further search for pulsations.  Similarly to how acceleration searches assume constant acceleration during an observation, jerk searches assume constant jerk with linearly varying acceleration \citep{Andersen2018}.  The short period nature of \jzeronine\ and \jtwentytwo\ warrant such additional searches.  We again used \presto\ and performed a jerk search with a \texttt{wmax} value of 50 for the first scan (MJD=59613) of \jzeronine\ and 200 for the second scan of \jzeronine\ and the scan of \jtwentytwo.  We visually inspected the preliminary folds for the time-series data for the top 100 candidates to reject RFI and null signals.  After folding the remaining candidates on the full time-series data, we still found no candidates that we believe are pulsar signals.
%find no parameter sets to claim as a detection of candidate detection.

\section{Future Outlook}\label{sec:outlook}
Although our observations do not produce a radio pulsation detection, they are by no means comprehensive.  We obtained our observations in a manner agnostic to orbital phase.  While detections of pulsations in spiders are not always limited to certain orbital phases, certain characteristics of the binary can limit our ability to detect pulsations.  Radio eclipses from circumbinary material can mask pulsations, and for RBs this masking can even happen for 25--100\% of the orbit.  Observations most sensitive to pulsations would occur at inferior conjunction, where heating of the companion would be at a maximum and material that can obscure pulsations would be most transparent.  

A possible spinning neutron star counterpart can never truly be ruled out; however, \citet{Ray2013} showed that in some cases it takes a substantial effort to verify the presence of a spider MSP.  Additional observations for each binary focused at inferior conjunction could provide more sensitive limits for a detection, though.  Continued monitoring may be necessary, especially in the case of \jzeronine, to rule out the scenario that the binaries are transitional MSPs and currently in a low-mass X-ray binary state.  The population of spider pulsars is still relatively small, so characterizing the nature of these candidate systems is a worthwhile pursuit.


% \begin{acknowledgments}
\section{Acknowledgments} \label{sec:acknowledgments}
The Green Bank Observatory is a facility of the National Science Foundation operated under cooperative agreement by Associated Universities, Inc. GBT observations were taken under project AGBT22A-355.
The National Radio Astronomy Observatory is a facility of the National Science Foundation operated under cooperative agreement by Associated Universities, Inc. KAC is supported by JWST-GO-02204.002-A.  Support for program \#2204 was provided by NASA through a grant from the Space Telescope Science Institute, which is operated by the Association of Universities for Research in Astronomy, Inc., under NASA contract NAS 5-03127. SMR is a CIFAR Fellow and is supported by the NSF Physics Frontiers Center awards 2020265.
% \end{acknowledgments}


\vspace{5mm}
\facilities{Robert C. Byrd Green Bank Telescope (GBT)}


\software{\presto\ \citep{PRESTO}, 
          \texttt{psrfits\_utils} 
          }

\bibliography{references}{}
\bibliographystyle{aasjournal}



\end{document}